\documentclass[debug]{rmaa}

\usepackage{mathtools}
\usepackage{paralist}

\usepackage{psfrag,color}

\usepackage[latin1]{inputenc}
\usepackage{float}
\usepackage[normalem]{ulem}

\usepackage{array}
\newcolumntype{P}[1]{>{\centering\arraybackslash}p{#1}}

\title{Dynamics Around an Asteroid Modeled as a Mass Tripole} 

\author{
  L. B. T. Santos,\altaffilmark{1} 
  L. O. Marchi,\altaffilmark{1}
  P. A. Sousa-Silva,\altaffilmark{2} 
  D. M. Sanchez,\altaffilmark{1} 
  S. Aljbaae,\altaffilmark{1}
  and A. F. B. A. Prado \altaffilmark{1}}

\altaffiltext{1}{National Institute for Space Research, INPE.}
\altaffiltext{2}{S\~ao Paulo University (UNESP) S\~ao Jo\~ao da Boa Vista}

\shortauthor{Santos, Marchi, Sousa-Silva, Sanchez, Aljbaae \& Prado}
\shorttitle{tripole model}

\fulladdresses{
\item Priscilla A. de Sousa-Silva: Department of Aeronautical Engineering - S\~ao Paulo University (UNESP), CEP
13876-750, Campus of S\~ao Jo\~ao da Boa Vista - SP, Brazil (priscilla.silva@unesp.br).
\item L. Barbosa T dos Santos: Division of Space Mechanics and Control - National Institute for Space Research
(INPE), CEP 12227-010, S~ao Jos'e dos Campos - SP, Brazil (leonardo.btorres@inpe.br). 
}

\listofauthors{L. B. T. dos Santos, L. Marchi, P. A. Sousa-Silva, D. M. Sanchez, S. Aljbaae  \& A. F. B. A. Prado}
\indexauthor{dos Santos, L. B. T.}
\indexauthor{Marchi, L.}
\indexauthor{Sousa-Silva, P. A.}
\indexauthor{Sanchez, D. M.}
\indexauthor{Aljbaae, S.}
\indexauthor{Prado, A. F. B. A.}

\abstract{The orbital dynamics of a spacecraft orbiting around irregular small celestial bodies is a challenging problem. Difficulties to model the gravity field of these bodies arise from the poor knowledge of the exact shape as observed from the Earth. In order to understand the complex dynamical environment in the vicinity of irregular asteroids, several studies have been conducted using simplified models. In this work, we investigate the qualitative dynamics in the vicinity of an asteroid with an arched shape using a tripole model based on the existence of three mass points linked to each other by rods with given lengths and negligible masses. We applied our results to some real systems, namely, asteroids 8567, 243 Ida and 433 Eros and also Phobos, one of the natural satellites of Mars.}

\resumen{}

\addkeyword{Methods: numerical}
\addkeyword{Minor planets}
\addkeyword{Asteroids: general space vehicles}

\begin{document}
% Typeset article header
\maketitle

\section{Introduction} 

Small-body explorations, such as asteroids and comets, have become an essential subject in deep space exploration. They involve multiple disciplines, such as science and control engineering, aerospace science and technology, celestial mechanics, astronomy among others. The combination of non-spherical gravitational attraction together with the rapid rotation of the asteroids around their axis govern the dynamics of the spacecraft near its surface. Thus, the analysis of the orbits of a spacecraft around these bodies is one of the current challenges in astrodynamics.

Developing mathematical models to represent the gravitational field around irregular bodies is an important research topic in orbital dynamics. Usually, spherical harmonics expansion is used to model the Earth and other Planets, as these more massive celestial bodies (when compared to asteroids) have a shape that resembles a sphere \citep{Elipe2002}. On the other hand, when the body does not resemble a sphere, this expansion is no longer convenient and, in some cases, convergence cannot be guaranteed \citep{Elipe2002}. Generally, when the field point is located within the circumscribing sphere, the series diverge \citep{2017Ap&SS.362..169L, Elipe2002}. Furthermore, the expansion of low-order Legendre coefficients often does not provide a good approximation for the motion of a spacecraft due to the fact that higher order terms can generate divergence after several iterations \citep{1999imda.coll..169R, 2018AdSpR..62.3199J}.

The shape of a celestial body, its rotation period, and other physical characteristics
can be obtained by light curve and radar analysis. From these observations, it is possible to use the solid polyhedron method to determine the dynamics around irregular bodies, including gravitational fields, stationary state solutions (equilibrium points, periodic orbits, quasiperiodic orbits, and chaotic motion), stability, bifurcation, etc \citep{werner, 1996Icar..121...67S, 2018AdSpR..62.3199J, chanut, 2014Ap&SS.349...83J, 2012AJ....143...62Y, Tsoulis}. However, this approach requires large computational effort depending on the quantity of polyhedral shapes. This problem was partially solved in  \citet{2015MNRAS.450.3742C}, where the authors considerably reduced the computation time ($\sim$ 30 times) applying the Mascon gravity framework, as presented in \citet{Geissler}, using a shaped polyhedral source to model the external gravitational field of a small celestial body. For more details about
this approach, we also refer the readers to \citet{Venditti} and \citet{2017MNRAS.464.3552A}.

The gravitational potential can be obtained with high accuracy using the polyhedral model, but from this model it is difficult to understand the effect of certain parameters (mass ratio ($\mu$), shape, among others) on the dynamics. This happens because, in the polyhedron model, the parameters mix and produce a mixed influence on the gravitational field of irregular bodies. Therefore, to study the effect of a single parameter, it is often necessary to model irregular bodies using simplified models.

By using simplified models, it is possible to perform semi analytical studies to understand which parameters affect stability, appearance of equilibrium points, bifurcations, etc. Thus, simplified models help to understand the dynamics around irregular bodies, and allow us to design orbits \citep{2017Ap&SS.362.9W, 2017ITAES..53.1221Z}, feedback control schemes \citep{2017Ap&SS.362...27Y}, as well as the permissible hovering regions \citep{2016JGCD...39.1223Z}.

An effective way to analyze the surface of an asteroid is to body-fixed hovering in a region close to the asteroide, where the spacecraft maintains its position constant with respect to the asteroid \citep{2020JGCD...43.1269W}. A great location for using the body-fixed hovering are the equilibrium points, due to the fact that they are locations that receive minimal disturbance. \citet{2014Ap&SS.349...83J} investigated body-fixed hovering at equilibrium points and classified the manifolds close to these points into eight types. Body-fixed hovering can be used to obtain accurate measurements of a region on the surface of the target asteroid and to facilitate the descent and ascend maneuvers of a spacecraft whose mission is to return to Earth with samples \citep{2005JGCD...28..343B}. Such maneuvers were used in the Hayabusa mission \citep{2004}.

Several bodies with different shapes can be described using simplified mathematical models. For example, \citet{Elipe2003, 1999imda.coll..169R, 2001CeMDA..81..235R}, analyze the motion of a particle under the gravitational field of a massive straight segment. A simple planar plate \citep{Blesa}, a rotating homogeneous cube \citep{2011Ap&SS.333..409L} and a triaxial ellipsoid \citep{2006SJADS...5..252G} have also been used to model bodies with irregular shapes.

\citet{2015Ap&SS.356...29Z} proposed that certain classes of elongated small bodies can be modeled by a double-particle-linkage called the dipole model. After that, \citet{2016Ap&SS.361...14Z} investigated the dynamical properties in the vicinity of an elongated body (using the dipole model) in order to analyze the influence of the force ratio ($k$), the mass ratio ($\mu$) and the oblateness ($A_2$) of the primary  in the distribution of the equilibrium points in the $xy$ plane. Through this dynamical analysis, \citet{2016Ap&SS.361...14Z} observed that the non-collinear equilibrium points exist only for 0.37 $\textless$ $k$ $\textless$ 2.07, and that these equilibria do not depend on $ \mu $. In \citet{2016Ap&SS.361...15Z}, the influence of parameters $k$, $\mu$ and $A_2$ (oblateness of the second primary) on the positions of the of out-of-plane equilibrium points and on the topological structure of the zero velocity curves were analyzed. \citet{2016Ap&SS.361...15Z} noted that the oblateness of the second primary greatly influences the distribution of equilibrium points outside the plane. These works, among others, showed that using that simplified model it is possible to identify the main parameters governing the dynamics around certain asteroid systems \citep{2017Ap&SS.362...61B, 2017Ap&SS.362..202D, 2018AJ....155...85Z}.

Inspired by the double-particle-linkage model, \citet{2017Ap&SS.362..169L} proposed that small arched bodies can be modeled by a triple-particle-linkage model determined by five parameters: $M$, $\omega$, $l_1$, $\tau$ and $\beta$. Analyzing asteroids 433 Eros, 243 Ida, and the Martian moon M1 Phobos, they validated the so called tripole model, by verifying that the gravitational field distribution of unstable annular regions is similar to the one found with the polyhedral model.
Later, \citet{2018RAA....18...84Y} proposed the non-axisymmetric triple particle-linkage model as a further step to improve the modeling towards a more realistic scenario. The authors analyzed the non-axisymmetric tripole model using three different elongated asteroids (243 Ida, 433 Eros, and (8567) 1996 HW1) and verified that the asymmetrical tripole model is more accurate than its predecessors, the dipole and the symmetrical tripole model.

We consider different geometries for the tripole to compute the gravitational potential and we compute the positions of the equilibrium points for the different combinations of relevant parameters of the model. Additionally, we analyze the conditions for linear stability. We find that the existence of some equilibrium points depends on the azimuthal angle and that the stability conditions depend on the rotation of the asteroids around their axis ($k$), on the azimuthal angle ($\Phi $), and on the mass ratio of the system ($\mu^*$). Also, we investigate the influence of $\Phi $ on the topological structure of the zero velocity curves. Finally, we find the relationship between the Jacobi constant and the azimuthal angle of the asteroid for all equilibrium points outside the asteroid's body.

Although the works found in the literature deal with the validation of the symmetric and the asymmetric tripole model, a semi-analytical analysis of the tripole model has not yet been performed. So, The main goal of the present work is to perform a dynamical analysis around arched asteroids and investigate which parameters ($k$, $\mu^*$ and $\Phi$, where $\Phi$ determines the degree of arching of the asteroid) influence in the distribution of the equilibrium points, in the topological structure of the zero velocity curves as well as the stability condition of stationary solutions. The tripole model has additional degrees of freedom when compared to the dipole model. So, it is possible to identify new parameters, such as the azimuthal angle, and to investigate their influence on the dynamical properties around an arched system.
With this, the results can be applied to investigate elongated natural arched bodies, such as some asteroid systems, comet nuclei and planet's moons.

We note that, from a dynamical point of view it should be interesting to explore the effect of the shape on the inner equilibria also. However, since we focus on the applicability of the solutions, we restrict the investigation to the points outside the body of the asteroid.

This article is organized as follows. The model and the methodology are discussed in Section \ref{methodology}. The results are analyzed and discussed in Section \ref{Results}. In section \ref{Application}, we investigate and compare the stability conditions of the model adopted in this study with real systems of small bodies. In section \ref{Conclusion}, some final considerations are made. 

\section{Mathematical Framework}
\label{methodology}

\begin{figure}[!t]
	\centering\includegraphics[scale=0.5]{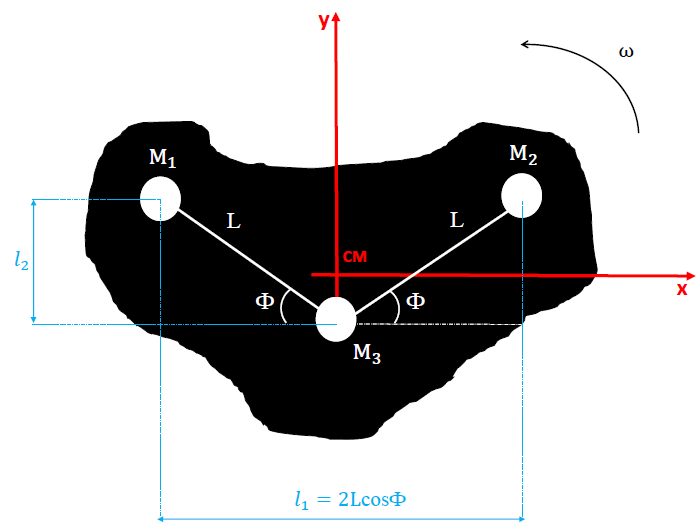}
	\caption{Schematic representation of the asteroid modeled by a tripole.}
	\label{fig1}
\end{figure}

In this section, we describe the Restricted Four-Body Problem using the rotating mass tripole model. In our investigations, we use the rotating mass tripole model shown in Fig. \ref{fig1}. This model consists in three mass points, $M_1$, $M_2$, and $M_3$, arranged inside an irregularly shaped asteroid. All the equations developed in this work refer to the asteroid-particle system (where particle is a body with negligible mass), i.e., the perturbations from other bodies are not taken into account.
The rods connecting $M_1$ to $M_3$ and $M_2$ to $M_3$ have negligible mass and the same length $L = 1$, which is the canonical unit. The distance between $M_1$ and $M_2$ is denoted by $l_1$, while the distance between $M_2$ and the $x$-axis, which contains $M_3$, is denoted by $l_2$. The parameter $\tau$ is defined as the ratio of $l_2$ to $l_1^*$, where $l_1^*$ = $l_1/2$, i.e. $\tau = l_2/l_1^*$. 

The origin of the reference system ($xy$) is at the center of mass of the asteroid. The angle formed by each rod with the $x$-axis is called the azimuthal angle and is denoted by $\Phi$ . We assume that both rods make the same angle  with the horizontal axis. The geometric configuration of the asteroid depends on this angle. The more arched the shape of the asteroid, the larger is the azimuthal angle. Note that when $\Phi = 0^{\circ}$ the length of the asteroid is maximum and equals to two canonical units. The equations that describe the motion of the particle in the $xy$ plane around the tripole are written in the rotating frame that rotates with constant angular velocity $\omega$ = 1, in canonical units. The unit of time is defined such that the period of rotation of the tripole is equal to 2$\pi$. We consider that $M_1$, $M_2$, and $M_3$ have equal masses, i.e., $m_1 = m_2 = m_3$.

\subsection{Equations of Motion}

Consider that the body with negligible mass (particle) is located at $P$($x$,$y$) and its motion is governed exclusively by the gravitational forces due to the primary bodies $M_1$, $M_2$, and $M_3$. $M_1$ and $M_2$ have masses $m_1$ = $m_2$ = $\mu^*$, and $M_3$ has mass $m_3 = 1-2\mu^*$, where $\mu^*$ is mass ratio defined as
\begin{equation}
\label{massratio}
\mu^* = \frac {m_2} { m_1 + m_2 + m_3 } 
\end{equation}

The coordinates of the primaries, in canonical units, are, respectively, given by:  
\begin{equation}
\label{x1}
x1 = -cos(\Phi), ~~~ y_1 = sin(\Phi) - 2\mu^*sin(\Phi), ~~~z_1 = 0 
\end{equation}

\begin{equation}
\label{x2}
x2 = cos(\Phi), ~~~ y_2 = sin(\Phi) - 2\mu^*sin(\Phi), ~~~z_2 = 0 
\end{equation}

\begin{equation}
\label{x3}
x_3 = 0, ~~~ y_3 =  - 2\mu^*sin(\Phi), ~~~z_3 = 0 
\end{equation}

Using the canonical units mentioned above the Hamilton function of the system is written as \citep{1968port.book.....B}:
\begin{equation}
\label{H}
H =  \frac {(p_x+y)^2 + (p_y + x)^2} {2} -  \frac {x^2 + y^2} {2} -  k\left (\frac {\mu^*} {r_1}+\frac {\mu^*} {r_2}+\frac {1-2\mu^*} {r_3}\right )
\end{equation}
where
\begin{equation}
\label{r1}
r_1 = \sqrt{(x-x_1)^2 + (y-y_1)^2 + z^2},
\end{equation}
\begin{equation}
\label{r2}
r_2 = \sqrt{(x-x_2)^2 + (y-y_2)^2 + z^2},
\end{equation}
\begin{equation}
\label{r3}
r_3 = \sqrt{(x-x_3)^2 + (y-y_3)^2 + z^2},
\end{equation}
and $p_x$ and $p_y$ are the components of the angular momentum of the particle with respect to the $x$-axis and the $y$-axis, respectively. The dimensionless parameter $k$ is the force ratio, given by the ratio between the gravitational force and the centrifugal force \citep{2018AJ....155...85Z, 2016Ap&SS.361...14Z} given by
\begin{equation}
\label{k}
k = \frac{G^*M}{\omega^{*2}l_1^{*3}}
\end{equation}

The value of $k$ depends on the angular velocity of the asteroid ($\omega^*$) in the international system of units, the total mass of the body ($M$) in kg and the length $l_1^*$ is the distance between $M_1$ and $M_2$ , in meters, and  $G^*$ is the universal gravitational constant in the international unit system. So $k$ can be computed after obtaining the length of the segment $l_1^*$. \citep{2018AJ....155...85Z,2017Ap&SS.362..169L, 2016Ap&SS.361...15Z} 

From the Hamilton function, it is possible to obtain the equations of motion of the particle in the rotating reference system:
\begin{equation}
\label{xdot}
\dot{x} = \frac{\partial H}{\partial p_x} = p_x + y
\end{equation}
\begin{equation}
\label{ydot}
\dot{y} = \frac{\partial H}{\partial p_y} = p_y - x
\end{equation}

The remaining dynamical equations are
\begin{equation}
\label{pxdot}
\dot{p}_x = -\frac{\partial H}{\partial x} = p_y - x + \Omega_x,
\end{equation}

\begin{equation}
\label{pydot}
\dot{p}_y = -\frac{\partial H}{\partial y} = p_x - y + \Omega_y,
\end{equation}
where $\Omega_x$ and $\Omega_y$ is the partial derivatives of $\Omega$ with respect to $x$ and $y$, respectively, that given by
\begin{equation}
\label{Omega}
\Omega = \frac {x^2 + y^2} {2} + k\left (\frac {\mu^*} {r_1}+\frac {\mu^*} {r_2}+\frac {1-2\mu^*} {r_3}\right ),
\end{equation}
Equation \ref{Omega} is a scalar function, also known as the pseudo-potential, which accounts for the acceleration experienced by the particle in a non-inertial reference system. 
The equations of motion in $xy$ plane in the Lagrangian formulation are \citep{1967torp.book.....S, 1999ssd..book.....M, 1963icm..book.....M, 2012CeMDA.113..291S}: 
\begin{equation}
\label{xdotdot}
\ddot{x} - 2\dot{y} = \Omega_x,
\end{equation}
\begin{equation}
\label{ydotdot}
\ddot{y} + 2\dot{x} = \Omega_y,
\end{equation}
which have the same appearance as the equations of the Classical Restricted Three-Body Problem (CRTBP) \citep{1914icm..book.....M, 1967torp.book.....S, 1999ssd..book.....M, 1963icm..book.....M}. 

Considering the motion in the $xy$ plane and multiplying Eq. \ref{xdotdot} by 2$x$ and Eq. \ref{ydotdot} by 2$y$, and adding all of them, we have that
\begin{equation}
\label{soma}
2\dot{x}\ddot{x} + 2\dot{y}\ddot{y} = 2\dot{x}\frac{\partial \Omega}{\partial y} + 2\dot{y}\frac{\partial \Omega}{\partial y} 
\end{equation}
which can be rewritten as

\begin{equation}
\label{v}
\frac{\mathrm{d}(\dot{x}^2 + \dot{y}^2)}{\mathrm{d} t} = 2 \frac{\partial \Omega}{\partial t} 
\end{equation}
Integrating Eq. \ref{v} with respect to time, we find that
\begin{equation}
\label{v2}
v^2 = 2\Omega - C^*
\end{equation}
where $v$ is the velocity of the particle and $C^*$ is a constant of integration.

In this paper, $C^*$ is called the modified Jacobi constant, where modified means that it is different from the constant studied by Jacobi for the case of the Classical Restricted Three-Body Problem. A special case occurs when $k$ = 1, since the modified Jacobi constant has the same value as the Jacobi constant, corresponding to the CRTBP. Looking at Eq. \ref{v2}, we note that the velocity of the particle depends only on the pseudo-potential and the integration constant $C^*$. The constant $C^*$ is determined numerically in terms of the initial position and velocity of the particle.

\subsection{Equilibrium Points}

Equilibrium solutions are points in which the particle has zero acceleration and zero velocity in the rotating frame. They are good locations in space to insert the spacecraft because they are located in regions where external perturbations are minimal, reducing the fuel consumption required for station-keeping maneuvers \citep{2017Ap&SS.362...61B}. 
The locations of the equilibrium points are explicitly defined in terms of $\mu^*$ (and implicitly by $\Phi$). Making the right side of Eqs. \ref{xdotdot} and \ref{ydotdot} equal to zero, that is, $\dot{x} = \dot{y} = 0$, implies null accelerations: 

\begin{equation}\label{eqsystem}
\begin{split}
x - k\frac{\mu^*(x-x_1)}{[(x-x_1)^2 + (y-y_1)^2]^{\frac{3}{2}}}
- k\frac{\mu^*(x-x_2)}{[(x-x_2)^2 + (y-y_2)^2]^{\frac{3}{2}}} \\ 
- k\frac{(1-2\mu^*)(x-x_3)}{[(x-x_3)^2 + (y-y_3)^2]^{\frac{3}{2}}} = 0, \\
~
~
~
y - k\frac{\mu^*(y-y_1)}{[(x-x_1)^2 + (y-y_1)^2]^{\frac{3}{2}}}
- k\frac{\mu^*(y-y_2)}{[(x-x_2)^2 + (y-y_2)^2]^{\frac{3}{2}}} \\ 
- k\frac{(1-2\mu^*)(y-y_3)}{[(x-x_3)^2 + (y-y_3)^2]^{\frac{3}{2}}} = 0.
% 		\label{Lcopla}
\end{split}
\end{equation}

The solutions of this system of equations can be determined numerically using an iterative method.

\subsection{Linear Stability Analysis.}

The linear stability analysis of the equilibrium points ($x_0$,$y_0$) is performed by displacing the origin of the coordinate system to the position of the libration points so that the equations of motion are linearized around the origin. Equation \ref{xdotdot} and \ref{ydotdot} can be written as, respectively
\begin{equation}\label{Stably}
\begin{split}
\ddot{\xi} - 2\dot{\eta} &= \Omega_{xx}(x_0,y_0)\xi + \Omega_{xy}(x_0,y_0)\eta, \\
% 	\label{Stably1}
% \end{equation}
% \begin{equation}
\ddot{\eta} + 2\dot{\xi} &= \Omega_{xy}(x_0,y_0)\xi + \Omega_{yy}(x_0,y_0)\eta,
\end{split}
% 	\label{Stably2}
\end{equation}
where the partial derivatives in ($x_0$,$y_0$) means that the value is computed at the libration point that is being investigated. $\xi$ and $\eta$ represent the coordinates of the particle with respect to the equilibrium point ($x_0$, $y_0$), and $\Omega_{xx}$, $\Omega_{xy}$, $\Omega_{xy}$, and $\Omega_{yy}$ are the partial derivatives calculated at this point, given by
\begin{equation}
\begin{split}
\Omega_{xx} = k\frac{3 (1-2\mu^*) x^2}{(x^2+(y-y_3)^2)^{5/2}}-k\frac{1-2
	\mu^*}{(x^2+(y-y_3)^2)^{3/2}}-k\frac{\mu^*}{((x-x_1)^2+(y-y_1)^2)^{3/2}}\\+k\frac{3 \mu^*
	(x-x_1)^2}{((x-x_1)^2+(y-y_1)^2)^{5/2}}-k\frac{\mu^*}{((x-x_2)^2+(y-y_2)^2)^{3/2}}\\+k\frac{3\mu^*
	(x-x_2)^2}{((x-x_2)^2+(y-y_2)^2)^{5/2}}+1, \\
\\
\Omega_{yy}= k\frac{3 (1-2 \mu^*) (y-y_3)^2}{\left(x^2+(y-y_3)^2\right)^{5/2}}-k\frac{1-2
	\mu^*}{\left(x^2+(y-y_3)^2\right)^{3/2}}+k\frac{3 \mu^*
	(y-y_1)^2}{\left((x-x_1)^2+(y-y_1)^2\right)^{5/2}}\\-k\frac{\mu^*}{\left((x-x_1)^2+(y-y_1)^2\right)^{3/2}}+k\frac{3 \mu^*
	(y-y_2)^2}{\left((x-x_2)^2+(y-y_2)^2\right)^{5/2}}\\-k\frac{\mu^*}{\left((x-x_2)^2+(y-y_2)^2\right)^{3/2}}+1, \\
\\
\Omega_{xy} = \Omega_{yx} =k\frac{3 (1-2 \mu^*)x (y-y_3)}{\left(x^2+(y-y_3)^2\right)^{5/2}}+k\frac{3 \mu^*(x-x_1)(y-y_1)}{\left((x-x_1)^2+(y-y1)^2\right)^{5/2}}\\+k\frac{3 \mu^*(x-x_2) (y-y_2)}{\left((x-x_2)^2+(y-y_2)^2\right)^{5/2}}.
 		\label{Uxy}
\end{split}
\end{equation}

The nontrivial roots of Eq. \ref{Stably} are obtained from the solution of the characteristic equation of order four in $\lambda$:
\begin{equation}
\lambda^4 + (4-\Omega_{xx}^0 - \Omega_{yy}^0)\lambda^2 +  \Omega_{xx}^0\Omega_{yy}^0 - (\Omega_{xy}^0)^2 = 0.  
\label{rchara}
\end{equation}

In Equation \ref{rchara}, $\Omega_{xx}^0$, $\Omega_{xy}^0$ and $\Omega_{yy}^0$ refer, respectively, to 
$\Omega_{xx}(x_0,~y_0)$, $\Omega_{xy}(x_0,~y_0)$ and $\Omega_{yy}(x_0,~y_0)$. 
The equilibrium point is linearly stable if all the four roots (or eigenvalues $\lambda$) of Eq. \ref{rchara} are purely imaginary, or complex with negative real parts \citep{2004CeMDA..90...87O}. However, if one or more of the eigenvalues have a positive real part, the equilibrium point is classified as unstable \citep{1914icm..book.....M, 1967torp.book.....S, 1999ssd..book.....M, 1963icm..book.....M}.

\section{Results}
\label{Results}

In this section we will show the numerical results obtained from our numerical simulations. The goal is to get a general view of the dynamics of the problem, which will allow us to get some conclusions.

\subsection{Influence of [$k$, $\mu^*$, $\Phi$] on equilibrium points}

\begin{figure}[!t]\centering
	\hspace*{6.4cm} \textbf{(a)} \\
	\includegraphics[scale=0.8]{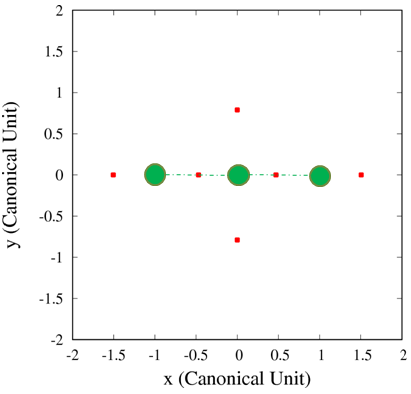} \\
	\hspace*{6.4cm} \textbf{(b)} \\
	\includegraphics[scale=0.8]{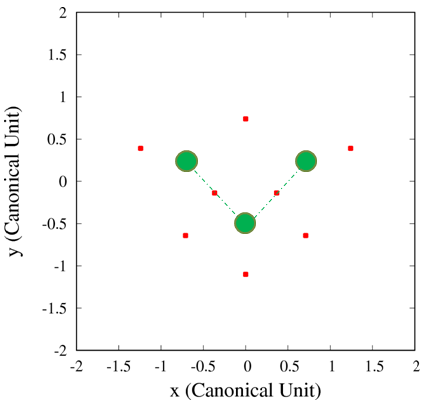}
	\caption{Equilibrium points for an azimuthal angle of (a) 0{\textdegree}
		and (b) 45{\textdegree}. In both cases $\mu^*=1/3$.}
	\label{Fig3}
\end{figure}

We start by computing the equilibrium points of the system. Figure \ref{Fig3}(a) shows the points of mass $M_1$ (green circle on the left side), $M_2$ (green circle on the right side) and $M_3$ (green middle circle), and six equilibrium points (red) for $\Phi = 0^{\circ}$, $\mu^*$ = 1/3 and $k$ = 1. The equilibrium points between $M_1$ and $M_3$ and between $M_2$ and $M_3$ overlap with the rod that connects the spheres (see Fig. \ref{fig1}). Therefore, we assume that these equilibrium points are inside the body of the asteroid.

Figure \ref{Fig3}(b) is similar to Fig. \ref{Fig3}(a), but now $\mu^*$ = 1/3 and $k$ = 1, but $\Phi$ = 45$^{\circ}$. In this case, there are eight equilibrium points, all of them off the $x$-axis. The position shift occurs because a new configuration is necessary to fulfill the equilibrium conditions as the positions of the primaries change, modifying the value of the azimuthal angle.

\begin{figure}[!t]
	\centering\includegraphics[scale=0.85]{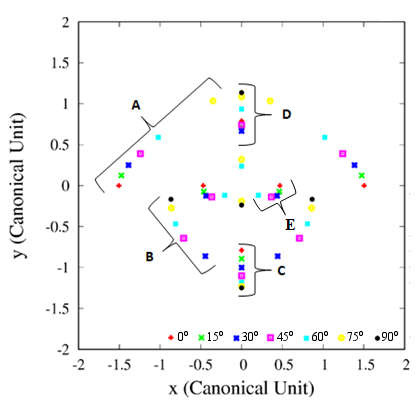}
	\caption{Equilibrium points separated by regions of A to E.}
	\label{Fig55}
\end{figure}

We performed numerical investigations to understand how the coordinates of the external equilibrium points change when $\mu^*$, $k$ and $\Phi$ are varied. To facilitate this analysis, we identified five regions, A, B, C, D, and E, as shown in Fig. \ref{Fig55}. We note that the regions are symmetric with respect to the $y$-axis. Observe that regions A, B and E are symmetric with respect to the $y$-axis, i.e., if the equilibrium point (in regions A, B or E) has coordinates ($x$, $y$), then there will be another equilibrium point in the coordinates (-$x$, $y$). Due to this symmetric property of the regions, we will only analyze the situations for which $x$ is negative. Figure \ref{Fig55} displays the equilibrium points when $\mu^*$ = 1/3, $k$ = 1 and $\Phi$ is varying. It illustrates how the equilibrium points move as this parameter is varied. The corresponding azimuthal angles are given in the caption of the plot. One can note the ``path" followed by the equilibrium points as $\Phi$ increases. For $\Phi$ = 90$^{\circ}$, the equilibrium points are equivalent to a dipole aligned at $x$ = 0.

\begin{figure}[!t]\centering
	\hspace*{6.4cm} \textbf{(a)} \\
	\includegraphics[scale=1.2]{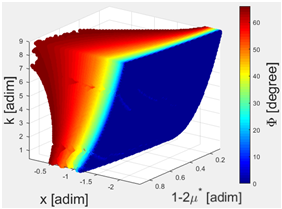} \\
	\hspace*{6.4cm} \textbf{(b)} \\
	\includegraphics[scale=1.2]{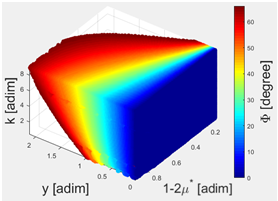}
	\caption{Coordinates of the equilibrium points in region A as a function of $k$, $\mu^*$ and $\Phi$. (\textit{a}) Position of the equilibrium points on $x$-axis. (\textit{b}) Position of equilibrium the points on $y$-axis.}
	\label{Aregion}
\end{figure}

For region A, we plot the behavior of the equilibrium points in the $x$ and $y$ plane as a function of $\mu^*$, $k$ and $\Phi$, as shown in Fig. \ref{Aregion}.

Figure \ref{Aregion} shows how the coordinates of the equilibrium points varies with  
$k$, $\mu^*$ and $\Phi$. Note that the graphs show the variation in the mass of the body $M_3$, given by 1-2$\mu^*$. That is, if the mass of $M_3$ increases, consequently the mass of $M_1$ (and $M_2$), given by $\mu^*$, decreases. The color bar represents the value of the azimuthal angle. First, we investigate the solutions when we vary $k$ and keep $\mu^*$ and $\Phi$ constant. Note from Fig. \ref{Aregion} that as the rotation of the asteroid decreases, that is, as $k$ becomes larger, the equilibrium points move away from the center of mass of the system. This is because increasing $k$ implies in decreasing the angular velocity of the asteroid around its own axis (see Eq. \ref{k}), thus making the value of the centrifugal force smaller. The condition of the existence of an equilibrium point is that the resulting force at one point in space is zero, that is, the gravitational force and the centrifugal force must have the same value in the module, but in the opposite direction. So, to keep the centrifugal force at a value that counteracts the gravitational force, the distance from the center of mass to the position of the equilibrium points increases. As $k$ increases, equilibrium points appear farther from the center of mass.

Next, keeping the values of $k$ and $\Phi$ constant and varying $\mu^*$, Fig. \ref{Aregion} shows that, as $\mu^*$ becomes smaller, the equilibrium points on the $x$-axis inside region A approach the center of mass of the system. On the other hand, as 1-2$\mu^*$ decrease (i.e. $\mu^*$ increase), the equilibrum points move away from asteroid. This happens because, as the mass of the bodies $M_1$ and $M_2$ becomes smaller, the gravitational force on the asteroid edge decreases on the $x$-axis, making it necessary to reduce the centrifugal force of the system on this axis. Conversely, the positions of the libration points to move away from the center of mass along the $y$-axis as the gravitational force on this axis becomes larger due to the increase in the mass of $M_3$. 

Finally, as we increase $\Phi$, the equilibrium points in region A along the $x$-axis come nearer to the center of mass of the system, while the ones along the $y$-axis move away from the center of mass of the system. These equilibrium points only exist when the azimuthal angle is between 0$^\circ$ and 76$^\circ$. Beyond this value, the configuration of the tripole does not allow the existence of equilibrium points in region A.

\begin{figure}[!t]\centering
		\hspace*{6.4cm} \textbf{(a)} \\
		\includegraphics[scale=0.6]{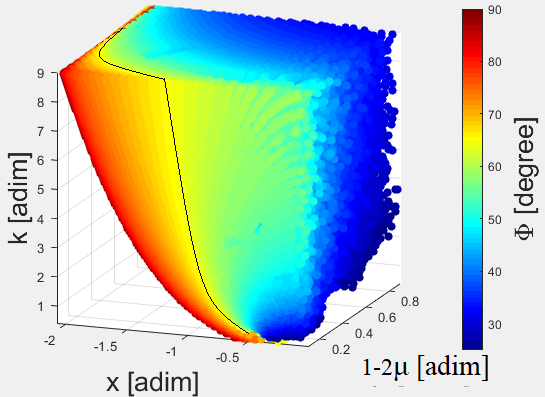} \\
		\hspace*{6.4cm} \textbf{(b)} \\
		\includegraphics[scale=0.6]{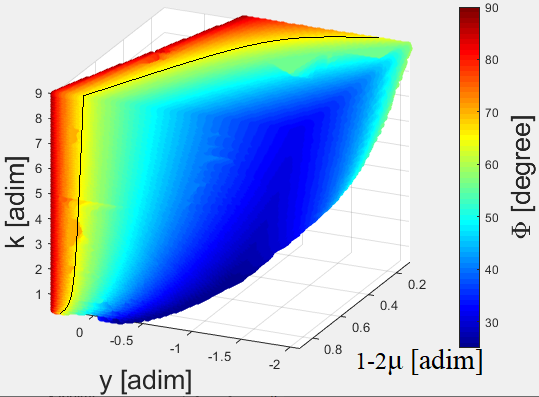}
		\caption{Coordinates of the equilibrium points in region B as a function of $k$, $\mu^*$ and $\Phi$. (\textit{a}) Position of the equilibrium points on $x$-axis. (\textit{b}) Position of equilibrium the points on $y$-axis.}
		\label{Bregion}
	\end{figure}

For region B, the variations of the $x$ and $y$ coordinates of the equilibrium points as a function of $k$, $\mu^*$ and $\Phi$  are shown in Figs. \ref{Bregion} (a) and (b), respectively. For a better view of the path taken by the equilibrium points when we vary the parameters $k$, $\mu^*$ and $\Phi$ , we insert a curve (black line) in the yellow region ($\Phi$ = 65$^\circ$). 

We note that, as $k$ becomes smaller, the positions of the equilibrium points in region B shift away from the center of mass of the system. This is true for the equilibrium points on both the $x$-axis and the $y$-axis and it occurs for the same reason as for the equilibrium points in region A.

The equilibrium points in region B occur for $\Phi$ $\textgreater$ 26$^\circ$. When the mass $m_3$ increases, the equilbirium points tend to move away from primary body on both the $x$ and $y$ axis. 

Analyzing the disposition of equilibrium points in region B as we increase $\Phi$, we find that the equilibrium points move upwards along the $y$-axis and may cross to the positive semi-plane. Unlike what happens for solutions in region A, the equilibrium points in region B move away from the system's center of mass along the $x$-axis increases.
	
Table \ref{tab1} summarizes the direction of the displacement of the equilibrium points in region A and B with respect to the body's center of mass as $k$, $\mu^*$ and $\Phi$  vary. The symbol $\nearrow$ indicates that the corresponding parameter is increasing, while $\equiv$ is used to indicate parameters that are fixed. Directional arrows denote the direction of the displacement of the equilibrium points. For example, when we keep fixed the values of $k$ and $\mu^*$, and increase $\Phi$, the equilibrium points of region A, $x$ and $y$, move to the right (approaching the system center of mass) and up (moving away from the system center of mass), respectively.

%	\begin{tabular}{ |P{3.2cm}|P{1.6cm}|P{1.6cm}|P{1.6cm}|P{1.6cm}| } 
%\hline
%\multicolumn{3}{|P{2.8cm}|} {Summary A region} & \multicolumn{3}{|P{2.8cm}|} {Summary A region} & %\multicolumn{2}{|P{2.8cm}|}{Summary B region} \\
%\hline
\begin{table}[!htbp]
	\small
	\begin{center}
		\centering
		\caption{Variation Trends of Coordinates for the Equilibrium points of the A and B regions}
		\label{tab1}
		\begin{tabular}{ |P{3.2cm}|P{1.6cm}|P{1.6cm}|P{1.6cm}|P{1.6cm}| } 
			\hline
			\multicolumn{1}{|P{3.2cm}|} {Summary, variation of parameters.} & \multicolumn{2}{|P{3.2cm}|} {Equilibrium point motion. A region} & \multicolumn{2}{|P{3.2cm}|}{Equilibrium point motion. B region} \\
			\hline
		    	& $x_0$ & $y_0$ & $x_0$ & $y_0$ \\ 
			\hline
			$k$ $\nearrow$, 1-2$\mu^*$ $\equiv$, $\Phi$ $\equiv$ & $\leftarrow$ & $\uparrow$ & $\leftarrow$ & $\downarrow$\\ 
			$k$ $\equiv$, 1-2$\mu^*$ $\nearrow$, $\Phi$ $\equiv$ & $\rightarrow$ & $\uparrow$  & $\leftarrow$ & $\uparrow$\\
			$k$ $\equiv$, 1-2$\mu^*$ $\equiv$, $\Phi$ $\nearrow$ &$\rightarrow$ & $\uparrow$  & $\leftarrow$ & $\uparrow$ \\ 
			
			\hline
			
		\end{tabular}		
	\end{center}
\end{table}

Next, we investigate regions C and D. In these two regions, the coordinates of the equilibrium points on the $x$-axis is zero for all points. 

\begin{figure}[!t]
	\centering\includegraphics[scale=0.9]{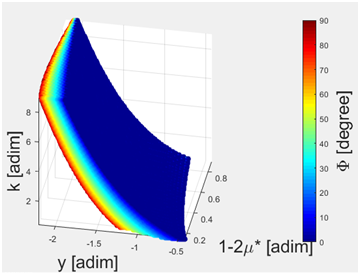}
	\caption{Behavior of the equilibrium points on $y$-axis of region C as a function of parameters $k$, $\mu^*$ and $\Phi$.}
	\label{Cregion}
\end{figure}

Figure \ref{Cregion} shows the $y$ coordinate of the equilibrium points as a function of $k$, $\mu^*$ and $\Phi$. When $k$ increases the centrifugal force becomes smaller, so the equilibrium points move downwards away from $M_3$. As the mass of $M_3$ increases, the gravitational force in the $y$ direction becomes stronger, causing the positions of the equilibrium points to change. As $\mu^*$ decreases and (1-2$\mu^*$) becomes larger, the equilibrium points in region C move in the negative direction of the $y$-axis.

Finally, as $\Phi$ increases, the equilibrium points move in the downwards along the $y$-axis. Figure \ref{Cregion} illustrates that the $y$ coordinate of the equilibrium points of the C region depends on $\Phi$, and that $y_C(\Phi$) becomes smaller as we increase the azimuthal angle. This happens because, as we increase $\Phi$, $M_1$ and $M_2$ move upwards along the $y$-axis. Then, to keep the center of mass of the system at the origin, $M_3$ must be in the semiplane with negative $y$-axis. Moreover, as $\Phi$ increases, the coordinate of $M_3$ becomes increasingly negative, so the equilibrium points in region C move away from $M_3$ in the negative direction to maintain the balance between the gravitational and centrifugal forces.

\begin{figure}[!t]
	\centering\includegraphics[scale=0.9]{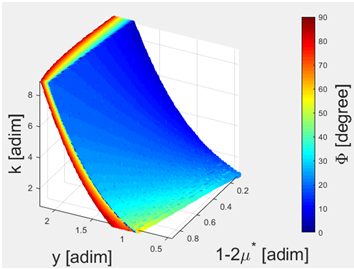}
	\caption{Behavior of the equilibrium points on $y$-axis of region D as a function of parameters $k$, $\mu^*$ and $\Phi$.}
	\label{Dregion}
\end{figure}

Figure \ref{Dregion} shows how the equilibrium points in region D depend on $k$, $\mu^*$ and $\Phi$. As $k$ increases, the equilibrium points move upwards away from the center of mass of the system. As the mass of $M_3$ increases, the gravitational force in the $y$ direction becomes larger, changing the positions of the equilibrium points. As (1-2$\mu^*$) becomes larger, the equilibrium points in region D move in the positive direction of the $y$-axis, away from the center of mass of the system.

\begin{figure}[!t]
	\centering\includegraphics[scale=1]{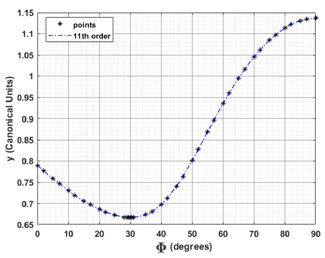}
	\caption{$y$-coordinate of the equilibrium points in region D as a function of $\Phi$ for [$\mu^*$, $k$] = [1/3, 1]$^T$. }
	\label{DDregion}
\end{figure}

Finally, we investigated the behavior of the equilibrium points on the $y$-axis when we increase $\Phi$. 
Initially, when we increase $\Phi$, the equilibrium points on the $y$-axis approaches the center of mass of the system. This happens because, as we increase the azimuthal angle, $M_3$ moves downward, consequently the gravitational force on the positive $y$-axis becomes weaker. In contrast, as we increase $\Phi$, the bodies $M_1$ and $M_2$ move upward with respect to the $y$-axis. This causes the gravitational force to increase in region D, now causing the equilibrium points to move upwards. For a better understanding, we constructed a figure using [$\mu^*$, $k$] = [1/3, 1]$^T$, which shows the equilibrium point behavior in the D region when we vary $\Phi$, as shown in Fig. \ref{DDregion}.

Then, as we increase the value of $\Phi$, the equilibrium point values in region D decrease, approaching the center of mass of the system, it reaches a minimum in $\Phi_D$ = 30.32$^\circ$ and the $y$ position of the D region that depends on $\Phi$ is $y$($\Phi)_{D-min}$ = 0.6664 then it increase again, moving away from the center of mass of the system.

Table \ref{tab4} summarizes the direction in the displacement of the equilibrium points in region D relative to the asteroid's center of mass when $k$, $\mu^*$ and $\Phi$ vary.
\begin{table}[!htbp]
	\small
	\begin{center}
		\centering
		\caption{Variation Trends of Coordinates for the Equilibrium points of the C and D regions}
		\label{tab4}
		\begin{tabular}{ |P{3.2cm}|P{1.6cm}|P{1.6cm}|P{1.6cm}|P{1.6cm}| } 
			\hline
			\multicolumn{1}{|P{3.2cm}|} {Summary, variation of parameters.} & \multicolumn{2}{|P{3.2cm}|} {Equilibrium point motion. C region} & \multicolumn{2}{|P{3.2cm}|}{Equilibrium point motion. D region} \\
			\hline
			& $x_0$ & $y_0$ & $x_0$ & $y_0$ \\ 
			\hline
			$k$ $\nearrow$, 1-2$\mu^*$ $\equiv$, $\Phi$ $\equiv$ & $0$ &  $\downarrow$ & $0$ & $\uparrow$\\ 
			$k$ $\equiv$, 1-2$\mu^*$ $\nearrow$, $\Phi$ $\equiv$ & $0$&  $\downarrow$  & $0$ & $\uparrow$\\
			$k$ $\equiv$, 1-2$\mu^*$ $\equiv$, $\Phi$ $\nearrow$ & $0$ &  $\downarrow$  & $0$ & $\downarrow$~$\uparrow$ \\ 
			
			\hline
			
		\end{tabular}	
	\end{center}
\end{table}

\subsection{Influence of azimutal angle on zero velocity curves }

The azimuthal angle is one of the mains parameters that govern the topological structure of the zero velocity curves around the tripole system. In this section, this effect is investigated. For the numerical simulations we keep [$k$, $\mu^*$]$^T$ = [1, 1/3]$^T$ and we vary the angle $\Phi$ in the interval [0, 90$^\circ$].

Equation \ref{v2} relates the square of the velocity and the position of the infinitesimal mass body in a rotating coordinate system. Note that when the integration constant $C^*$ is numerically determined by the initial conditions, Equation \ref{v2} gives the speed with which the infinitesimal mass body moves.

\begin{figure*}[!h]
	\newlength\thisfigwidth
	\setlength\thisfigwidth{0.5\linewidth}
	\addtolength\thisfigwidth{-0.5cm}
	
	\makebox[\thisfigwidth][l]{\textbf{a}} 
	\hfill%
	\makebox[\thisfigwidth][l]{\textbf{b}}\\[-3ex]
	\parbox[t]{\linewidth}{%
		\vspace{0pt}
		\includegraphics[width=6cm,height=6cm]{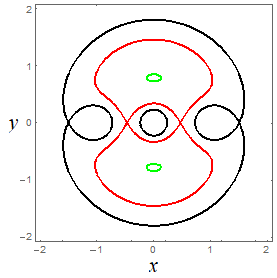}%
		\hfill%
		\includegraphics[width=6cm,height=6cm]{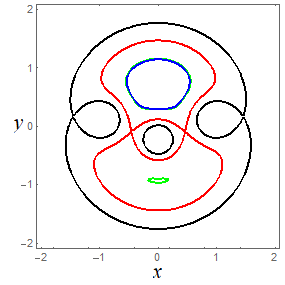}}
	~
	~
	
	\makebox[\thisfigwidth][l]{\textbf{c}}
	\hfill%
	\makebox[\thisfigwidth][l]{\textbf{d}}\\[-3ex]
	\parbox[t]{\linewidth}{
		\vspace{0pt}
		\includegraphics[width=6cm,height=6cm]{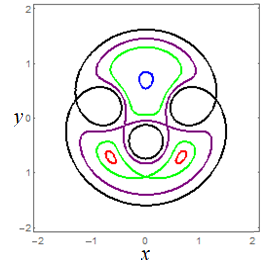}%
		\hfill%
		\includegraphics[width=6cm,height=6cm]{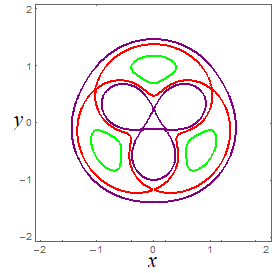}}
	~
	~
	
	\makebox[\thisfigwidth][l]{\textbf{e}}
	\hfill%
	\makebox[\thisfigwidth][l]{\textbf{f}}\\[-3ex]
	\parbox[t]{\linewidth}{
		\vspace{0pt}
		\includegraphics[width=6cm,height=6cm]{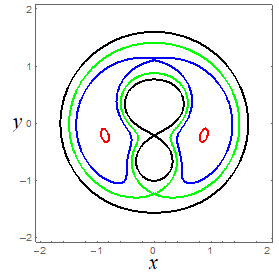}%
		\hfill%
		\includegraphics[width=6cm,height=6cm]{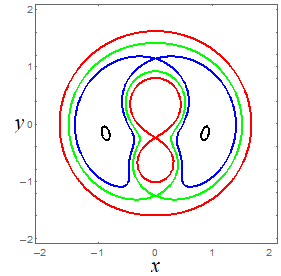}
	}
	\caption{Influence of the azimuthal angle on zero-velocity curves in the $xy$ plane. (\textit{a})	Zero velocity curves for a 0$^\circ$ azimuthal angle.
		(\textit{b})	Zero velocity curves for a 20$^\circ$ azimuthal angle. (\textit{c}) Zero velocity curves for a 40$^\circ$ azimuthal angle. (\textit{d}) Zero velocity curves for a 60$^\circ$ azimuthal angle. (\textit{e})	Zero velocity curves for a 80$^\circ$ azimuthal angle.
		(\textit{f})	Zero velocity curves for a 90$^\circ$ azimuthal angle.}
	\label{CVZ}
\end{figure*}

In particular, if $v$ is assigned zero, Equation \ref{v2} defines the curves at which the velocity is zero.
The equation that gives the zero velocity curves, in cartesian coordinates, is:
\begin{equation}
\label{v0}
x^2+y^2 + \frac{2\mu^*}{r_1} + \frac{2\mu^*}{r_2} + \frac{2(1-2\mu^*)}{r_3} = C^*
\end{equation}
where $r_1$, $r_2$ and $r_3$ are as shown in equations \ref{r1}, \ref{r2} and \ref{r3}.
The zero velocity curves in the $xy$ plane for six different values of $\Phi$ are shown in Fig. \ref{CVZ}. Each curve in frames a) to f) of Fig. \ref{CVZ} corresponds to the value of the Jacobi constant for which the contacts between the ovals occur and the equilibrium points appear. The tripole is not illustrated in the figure. 

Figure \ref{CVZ} a) shows the zero velocity curves when the azimuthal angle is 0$^\circ$. Note that, for this azimuthal angle, $M_1$, $M_2$, and $M_3$ are aligned on the $x$-axis. On the other hand, Figure \ref{CVZ} b) shows the zero velocity curves when the azimuthal angle is 20$^\circ$. For small values of $x$ and $y$ that satisfies Eq. \ref{v0}, the first two terms are virtually irrelevant and the equation can be written as:
\begin{equation}
\label{e0}
-\frac{\mu^*}{r_1} - \frac{\mu^*}{r_2} -\frac{2(1-2\mu^*)}{r_3} = \frac{C^*}{2} - \frac{(x^2+y^2)}{2} = \frac{C^*}{2} - \epsilon.
\end{equation}

This equation gives the equipotential curves for the three centers of force $\mu^*$, $\mu^*$ and 1-2$\mu^*$, as shown in Fig. \ref{CVZ} a) and b). For large values of $C^*$, ovals consist of closed curves around each of the body. If we decrease $C^*$, the ovals around $M_1$, $M_2$ and $M_3$ (inner ovals) expand, and the outer contours (outer ovals) move towards the center of mass of the system. The inner ovals connect with the outer ovals, resulting in the equilibrium points in region A (black curve) and the ovals between the bodies also connect, resulting in the equilibrium points in region E (red curve). See Figures \ref{CVZ} a) and b). 

If $C^*$ is further decreased, the regions where movement is allowed become larger. This happens because the oval around the masses increases and merges with the outer oval, leaving only a small confined area (regions C and D), where the movement is impossible. Note from Fig. \ref{CVZ} a) that, due to the symmetry of the problem, equilibrium points in regions C and D appear for the same value of $C^*$ (green curve). On the other hand, when the azimuthal angle is different from 0$^\circ$, the equilibrium points in the C and D regions appear for different Jacobi constant values (green and blue curves, respectively) shown in Fig. \ref{CVZ} b).

Figure \ref{CVZ} c) shows the zero velocity curves when the azimuthal angle is 40$^\circ$. The change in the topological structure of the zero velocity curves is evident as the azimuthal angle is varied. Note from Figure \ref{CVZ} c) that, in addition to the contact points shown in Figs. \ref{CVZ} a) and b), new contact points emerge (red curves), in region B. Through numerical simulations, we observe that the B regions arises when the azimuthal angle is greater than 26$^\circ$.

When we consider the azimuthal angle of 60$^\circ$, $M_1$, $M_2$ and $M_3$ form an equilateral triangle relative to the rotating reference system. 

\clearpage
Thus, the zero velocity curves has a symmetrical shape. When the azimuthal angle is 60$^\circ$, the equilibrium points in regions A and C arise for $C_{A-C}$ = 2.946725190. Likewise, the ($C_{B-D}$ = 3.35803516) is required for contacts between ovals in regions B and D. If the masses of $M_1$ and $M_2$ are different, the symmetrical property of the equilibrium points and zero velocity curves with respect to the $xy$ axis is not valid.
Figure \ref{CVZ} e) shows the zero velocity curves for an azimuthal angle of 80$^\circ$. In Figure \ref{CVZ} e), we observe that regions A cease to exist, leaving only regions B, C, D and E. This means that, just as regions B depend on the azimuthal angle to emerge or disappear, so does regions A.

Regions A ceases to exist for$\Phi$ ~ $\textgreater$ 76$^\circ$.

Finally, considering an azimuthal angle of 90$^\circ$, $M_1$ and $M_2$ overlap, which means that they behave as a single body with mass $m=m_1 + m_2$. For this configuration, the system is similar to the Classical Restricted Three-body Problem with a mass ratio of $\mu^* = 1/2$. 

Note from Figures \ref{CVZ} a) - f) that, as we increase the azimuthal angle from 0 to 90$^\circ$, noticeable changes in the zero velocity curves can be observed near the arched asteroid. Note that the regions that connect the ovals move along the $xy$ plane as we vary the azimuthal angle. Some fixed points also emerge or disappear.

\begin{figure*}[!ht]
	\setlength\thisfigwidth{0.5\linewidth}
	\addtolength\thisfigwidth{-0.5cm}
	
	\makebox[\thisfigwidth][l]{\textbf{a}} 
	\hfill%
	\makebox[\thisfigwidth][l]{\textbf{b}}\\[-3ex]
	\parbox[t]{\linewidth}{%
		\vspace{0pt}
		\includegraphics[width=6cm,height=6cm]{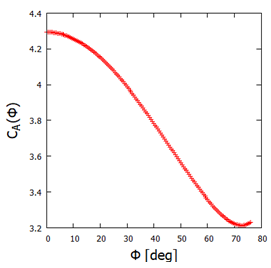}%
		\hfill%
		\includegraphics[width=6cm,height=6cm]{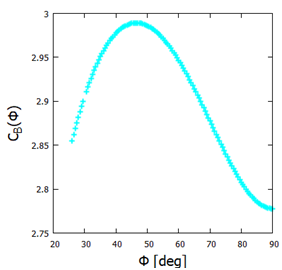}}
	~
	~
	
	\makebox[\thisfigwidth][l]{\textbf{c}}
	\hfill%
	\makebox[\thisfigwidth][l]{\textbf{d}}\\[-3ex]
	\parbox[t]{\linewidth}{
		\vspace{0pt}
		\includegraphics[width=6cm,height=6cm]{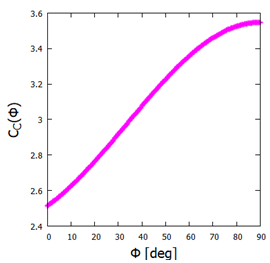}%
		\hfill%
		\includegraphics[width=6cm,height=6cm]{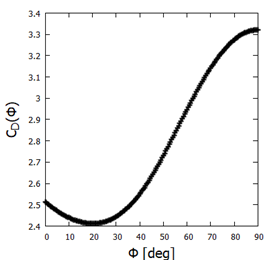}}

	\caption{Jacobi constant behavior in regions A, B, C and D, respectively, as a function of azimuthal angle. (\textit{a}) Values of the Jacobian constant ($C_A$) at the equilibrium points versus $\Phi$.
		(\textit{b})	Values of the Jacobian constant ($C_B$) at the equilibrium points versus $\Phi$. (\textit{c}) Values of the Jacobian constant ($C_C$) at the equilibrium points versus $\Phi$. (\textit{d}) Values of the Jacobian constant ($C_D$) at the equilibrium points versus $\Phi$.}
	\label{cbphi}
\end{figure*}

The values of the modified Jacobi constants at the contact points in each region in Fig. \ref{CVZ} are shown in Fig. \ref{cbphi}. Figures \ref{cbphi} a) - d) show how the values of the Jacobi constant at regions A, B, C, and D ($C_A$, $C_B$, $C_C$, and $C_D$, respectively) vary as a function of the azimuthal angle $\Phi$.
In Fig. \ref{cbphi} - a), we see that the values of the Jaccobi constant $C_A$($\Phi$) decrease as the azimuthal angle $\Phi$ increases. For $C_B$($\Phi$), one notes that, initially the value of the Jacobi constant increases with increasing azimuthal angle and then decreases, as shown in Fig. \ref{cbphi} b). This behavior causes a maximum value for $C_B$($\Phi$), which happens at $C_B$ = 2.989303755, for $\Phi$ = 46.524234$^\circ$. On the other hand, the values of the function $C_C$($\Phi$) increase as we increase $\Phi$. Finally, for $C_D$, as we increase $\Phi$, initially, the values of $C_D$ become smaller, reaching a minimum value of $C_D$ = 2.4120014, when the azimuthal angle is approximately $\Phi$ = 19.987$^\circ$, and then it increases.

\subsection{Stability conditions}

Now, we focus on the analysis of the stability conditions for the equilibrium points in regions D and C, ($L_D$ and $L_C$), respectively, i.e., points that have null $x$ coordinate. We describe how the stability conditions for the equilibrium points $L_D$ (and $L_C$) depend on the azimuthal angle ($\Phi$), the force ratio ($k$) and the mass ratio ($\mu^*$). Indeed, if any of these parameters are changed, the stability condition (unstable or stable) of these equilibrium points may also change. 

\begin{figure*}[!t]
	\setlength\thisfigwidth{0.5\linewidth}
	\addtolength\thisfigwidth{-0.5cm}
	
	\makebox[\thisfigwidth][l]{\textbf{a}} 
	\hfill%
	\makebox[\thisfigwidth][l]{\textbf{b}}\\[-3ex]
	\parbox[t]{\linewidth}{%
		\vspace{0pt}
		\includegraphics[width=6cm,height=6cm]{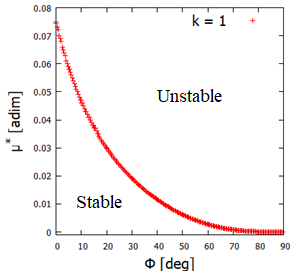}%
		\hfill%
		\includegraphics[width=6cm,height=6cm]{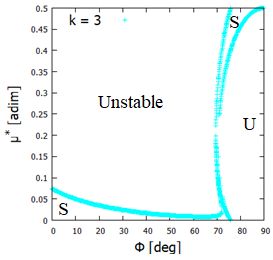}}
	~
	~
	
	\makebox[\thisfigwidth][l]{\textbf{c}}
	\hfill%
	\makebox[\thisfigwidth][l]{\textbf{d}}\\[-3ex]
	\parbox[t]{\linewidth}{
		\vspace{0pt}
		\includegraphics[width=6cm,height=6cm]{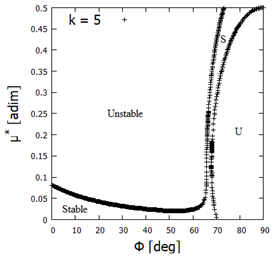}%
		\hfill%
		\includegraphics[width=6cm,height=6cm]{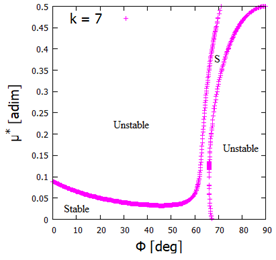}}
	
	\caption{Values of the mass ratio ($\mu^*$) versus the azimuthal angle ($\Phi$) for the stability condition of the equilibrium point $L_D$ considering different values of $k$.
	(\textit{a}) Values of the mass ratio ($\mu^*$) versus the azimuthal angle ($\Phi$) when $k$ = 1 for the stability condition of the equilibrium point $L_D$. 	(\textit{b})	Values of the mass ratio ($\mu^*$) versus the azimuthal angle ($\Phi$) when $k$ = 3 for the stability condition of the equilibrium point $L_D$. (\textit{c}) Values of the mass ratio ($\mu^*$) versus the azimuthal angle ($\Phi$) when $k$ = 5 for the stability condition of the equilibrium point $L_D$. (\textit{d}) Values of the mass ratio ($\mu^*$) versus the azimuthal angle ($\Phi$) when $k$ = 7 for the stability condition of the equilibrium point $L_D$.}
	\label{muphi}
\end{figure*}

First let's look at the stability condition for region D. Figure \ref{muphi} shows plots of $\Phi$ versus $\mu^*$, showing the stability transition. We see from Fig. \ref{muphi} a) that, when the azimuthal angle increases and $k = 1$, the mass ratio required to maintain the equilibrium point $L_D$ stable decreases. When the angle is 0$^\circ$, the maximum mass ratio to allow linear stability of the system studied is $\mu^*$ = 0.0742683. If the mass ratio is greater than this value, the system is unstable for every azimuthal angle. Note that, when $\Phi$ $\rightarrow$ 90$^\circ$, the two masses of the tripole ($m_1$ and $m_2$) colapse into a mass point with the mass ratio 2$\mu^*$. In this case, the point $L_D$ is similar to the equilibrium point $L_3$ of the Classical Restricted Three-Body Problem. Therefore this equilibrium point is linearly unstable for any mass ratio, which is in agreement with the literature \citep{1914icm..book.....M, 1967torp.book.....S, 1999ssd..book.....M, 1963icm..book.....M}.

Figures \ref{muphi} b) to d) show $\Phi$ versus $\mu^*$, which illustrate the stability regions when $k$ $\textgreater$ 1. We see from Fig. \ref{muphi} (b) that, for $\Phi$ $\textless$ 70$^\circ$, the stability transition is similar to the case when $k$ = 1, but the bifurcation occurs when $\Phi$ $\sim$ 70$^\circ$. Notice in the graph that a narrow vertical strip appears, causing the $L_D$ equilibrium point stable for any value of $\mu^*$. As $\Phi$ increases, the stability conditions change again, making the equilibrium point stable only for high values of $\mu^*$. So, observe that, when the system has low values of $\mu^*$, the equilibrium points are linearly stable for $\Phi$ $\textless$ 76$^\circ$. On the other hand, for a very arched asteroid ($\Phi$  $\textgreater$ 76$^\circ$), the equilibrium point $L_D$ is linearly stable when the mass ratio of the system is high. 

Figure \ref{muphi} c) shows the stability transition curve for k = 5. We observed that when $\Phi ~ < ~ 60^\circ$, the stability transition curve is similar to the previous cases. We also notice that narrow vertical strip appears (around $\Phi ~ \approx ~ 65^\circ$) and has a larger area with respect to the previous case. This means that we can also find stable regions when we consider high values of $\Phi$ ($\Phi ~ >  ~ 60^\circ$) and $\mu^*$. As we increase the value of $\Phi$ (when $\Phi ~ > 70^\circ$), the equilibrium point $L_D$ becomes linearly stable only for high values of $\mu^*$. For low values of $\mu^*$, the equilibrium point $L_D$ is stable when $\Phi ~ < ~ 70^\circ$. The letters S and U shown in Fig. \ref{muphi} c) are abbreviations for Stable and Unstable condition, respectively.

Finally, Fig. \ref{muphi} d) shows the stability transition when $k$ = 7. Note that, as in the previous cases, when we consider $k$ = 7, narrow vertical strip appears (around $\Phi ~~ 60^\circ$), allowing the equilibrium point $L_D$ to be linearly stable for any value of $\mu^*$. If we gradually increase $\mu$ and $\Phi$, the stable regions remain until $\Phi ~ = ~ 89.6^\circ$. On the other hand, if we decrease $\mu$ as we increase $\Phi$ (from $66^\circ$), the stable region extends to $\Phi ~ = ~ 67^\circ$. Note in Figures \ref{muphi} b) - d) that the area of the narrow vertical strip becomes larger as we increase the k value. This means that, the higher the value of k, the larger the region that allows linear stability of equilibrium point $L_D$ to any values of $\mu^*$.

\begin{figure*}[!t]
	\setlength\thisfigwidth{0.5\linewidth}
	\addtolength\thisfigwidth{-0.5cm}
	
	\makebox[\thisfigwidth][l]{\textbf{a}} 
	\hfill%
	\makebox[\thisfigwidth][l]{\textbf{b}}\\[-3ex]
	\parbox[t]{\linewidth}{%
		\vspace{0pt}
		\includegraphics[width=6cm,height=6cm]{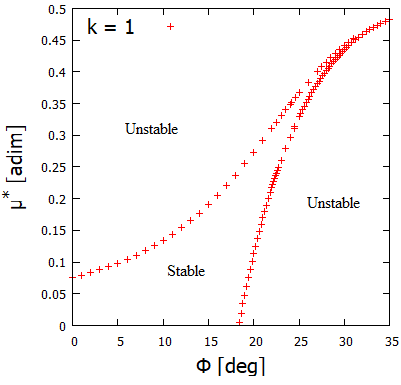}%
		\hfill%
		\includegraphics[width=6cm,height=6cm]{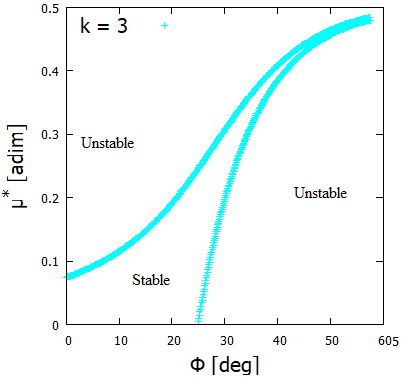}}
	~
	~
	
	\makebox[\thisfigwidth][l]{\textbf{c}}
	\hfill%
	\makebox[\thisfigwidth][l]{\textbf{d}}\\[-3ex]
	\parbox[t]{\linewidth}{
		\vspace{0pt}
		\includegraphics[width=6cm,height=6cm]{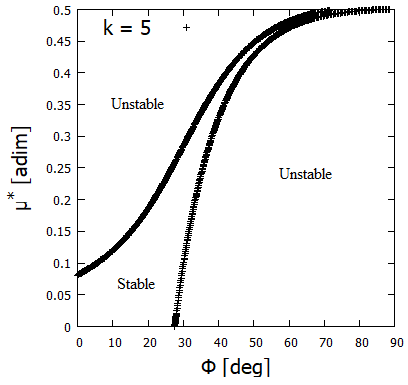}%
		\hfill%
		\includegraphics[width=6cm,height=6cm]{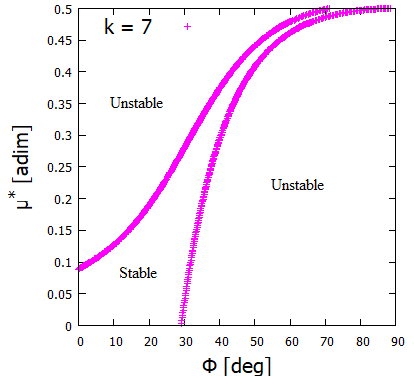}}
	
	\caption{Values of the mass ratio ($\mu^*$) versus the elevation angle ($\Phi$) for the stability condition of the equilibrium point $L_C$ considering different values of $k$. (\textit{a}) Values of the mass ratio ($\mu^*$) versus the azimuthal angle ($\Phi$) when $k$ = 1 for the stability condition of the equilibrium point $L_C$. 	(\textit{b})	Values of the mass ratio ($\mu^*$) versus the azimuthal angle ($\Phi$) when $k$ = 3 for the stability condition of the equilibrium point $L_C$. (\textit{c}) Values of the mass ratio ($\mu^*$) versus the azimuthal angle ($\Phi$) when $k$ = 5 for the stability condition of the equilibrium point $L_C$. (\textit{d}) Values of the mass ratio ($\mu^*$) versus the azimuthal angle ($\Phi$) when $k$ = 7 for the stability condition of the equilibrium point $L_C$.}
	\label{stable}
\end{figure*}

A similar analysis was performed for the equilibrium point $L_C$ and the results are shown in Fig. \ref{stable}.
Unlike Figure \ref{muphi} a), when $k$ = 1, Figure \ref{stable} a) shows that there are two stability transition limits. The first limit (lower transition, left-hand curve) exists for small azimuthal angles, starting at 0$^\circ$, with a mass ratio of 0.07427949. Above 18.351$^\circ$, numerical evidence shows that another stability transition arises, as shown by the right-hand curve in Fig. \ref{muphi} a). 

Figures \ref{stable} b) - d) show $\Phi$ versus $\mu^*$, which illustrates the stability regions, when $k$ $\textgreater$ 1. Figure \ref{stable} b) shows two stability transitions. Note that the first transition starts when $\Phi$ = 0$^\circ$ and $\mu^*$ is approximately 0.074.

The second stability transition starts when $\Phi$ = 25$^\circ$, when the asteroid is 8$\circ$ more arched than the previous case, so the equilibrium point $L_C$ has a wider stable region compared to when $k$ = 1. For $\Phi$ $\textgreater$ 57.5$^\circ$, the equilibrium point $L_C$ is unstable for any mass ratio.

If we further increase the value of $k$ to $k$ = 5, the stability region becomes even larger, as shown in Fig. \ref{stable} c). The first stability transition arises when $\Phi$ = 0$^\circ$ and $\mu^*$  = 0.08. In contrast, the second curve arises when $\mu^*$  = 0 and $\Phi$ = 28$^\circ$, thus limiting the region that allows the equilibrium point $L_C$ to be stable. If the azimuthal angle is greater than 68$^\circ$, the equilibrium point $L_C$ becomes unstable for any mass ratio.

Finally, we made an analysis considering $k$ = 7. Note from Figure \ref{stable} d) that, due to the low rotation of the asteroid, it results in a larger area on the graph that makes the $L_C$ equilibrium point linearly stable. For $k$ = 7, the first transition starts when $\Phi$ = 0$^\circ$ and $\mu^*$ = 0.09. In contrast, the second stability transition starts when $\Phi$ = 29$^\circ$ and $\mu^*$ = 0. This shows that, when we increase the value of $k$ (ie, the angular velocity of the asteroid becomes slower), the two stability transition curves intersect at a larger azimuthal angle, ranging from approximately, $\Phi$ = 35$^\circ$ when $k$ = 1, until $\Phi$ = 75$^\circ$ when $k$ = 7. This shows that, as we increase the force ratio $k$, the stability region becomes larger.

\section{Application}
\label{Application}

To validate the equations and results developed in this article, we compared the results obtained with four celestial bodies, (i) 243 Ida, (ii) 433 Eros, (iii) 1996(HW1) and (iv) M1 Phobos. 

The parameters $k$, $\Phi$ and $\mu^*$ were taken from \citet{2017Ap&SS.362..169L} (for Ida and M1 Phobos) and \citet{2018RAA....18...84Y} (for Eros and 1996 HW1). The linear stability of the equilibrium points of the celestial bodies mentioned above were obtained by \citet{2014Ap&SS.353..105W} and used in this study for comparison purposes. In \citet{2014Ap&SS.353..105W}, regions $C$ and $D$ are the equilibrium points $E_4$ and $E_2$, respectively. 

The optimized parameters of the bodies under analysis in this article are shown in Table \ref{parameterss},
\begin{table}[!t]\centering
	\setlength{\tabnotewidth}{0.5\columnwidth}
	\tablecols{4}
	% Stretch the space between table columns 
	\setlength{\tabcolsep}{2.8\tabcolsep}
	\caption{The optimal parameters for the tripole models} \label{parameterss}
	\begin{tabular}{lrrr}
		\hline
	    Asteroid &$k$ & $\mu^*$ & $\Phi$ \\
		\midrule
		243 Ida &0.402  &0.237 & 19.94$^\circ$\\
		M1 Phobos &22.003 & 0.396 &56.09$^\circ$\\
	    433 Eros&	0.434 &0.260 & 18.95$^\circ$\\
	    1996 (HW1)	&3.158 &0.443 & 27.43$^\circ$\\
		\hline
	\end{tabular}
\end{table}
where $\Phi$ is determined by doing $\Phi = \arctan{(2\sigma)}$ in which $\sigma$ is given by $l_2/l_1$ and was determined in \citet{2017Ap&SS.362..169L} and \citet{2018RAA....18...84Y}.

Knowing the parameters for each celestial body, it is possible to find the stability conditions for equilibrium points $E_4$ and $E_2$ from Equations \ref{Uxy} and \ref{rchara}.

\begin{figure*}[!ht]
	\setlength\thisfigwidth{0.5\linewidth}
	\addtolength\thisfigwidth{-0.5cm}
	
	\makebox[\thisfigwidth][l]{\textbf{a}} 
	\hfill%
	\makebox[\thisfigwidth][l]{\textbf{b}}\\[-3ex]
	\parbox[t]{\linewidth}{%
		\vspace{0pt}
		\includegraphics[width=6cm,height=6cm]{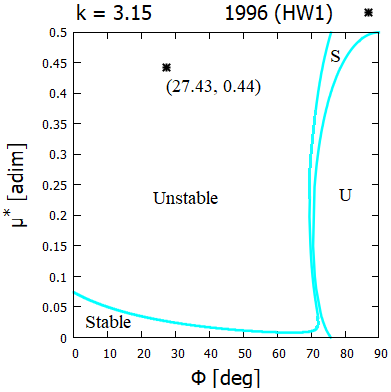}%
		\hfill%
		\includegraphics[width=6cm,height=6cm]{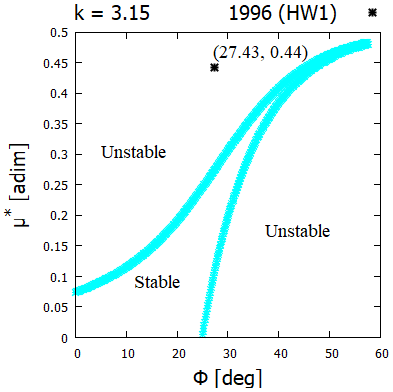}}
	~
	~
	
	\makebox[\thisfigwidth][l]{\textbf{c}}
	\hfill%
	\makebox[\thisfigwidth][l]{\textbf{d}}\\[-3ex]
	\parbox[t]{\linewidth}{
		\vspace{0pt}
		\includegraphics[width=6cm,height=6cm]{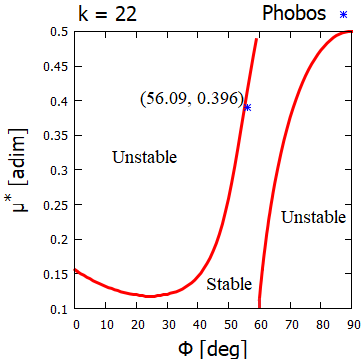}%
		\hfill%
		\includegraphics[width=6cm,height=6cm]{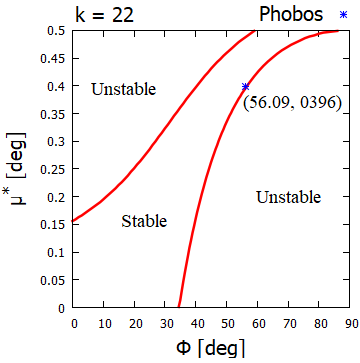}}
	~
	~
	
	\makebox[\thisfigwidth][l]{\textbf{e}}
	\hfill%
	\makebox[\thisfigwidth][l]{\textbf{f}}\\[-3ex]
	\parbox[t]{\linewidth}{
		\vspace{0pt}
		\includegraphics[width=6cm,height=6cm]{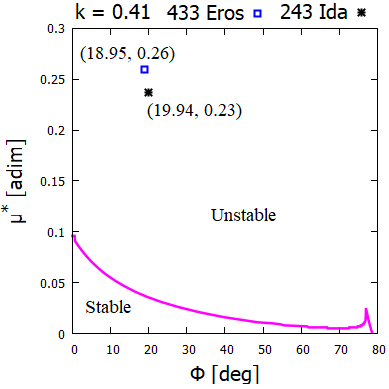}%
		\hfill%
		\includegraphics[width=6cm,height=6cm]{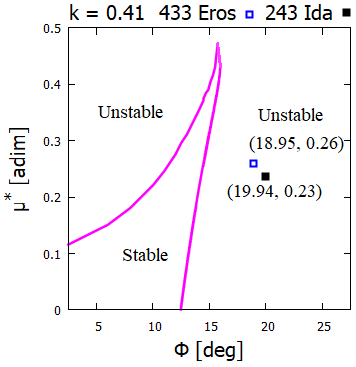}
	}
	\caption{Values of the $\mu^*$ $versus$ $\Phi$ for the stability condition of the equilibrium point $L_D$ ($E_2$) and $L_C$ ($E_4$) for a specific $k$ value.
		(\textit{a}) $k$ = 3.15 for the equilibrium point $L_D$ ($E_2$) of the 1996 HW1 asteroid. 	(\textit{b}) $k$ = 3.15 for the equilibrium point $L_C$ ($E_4$) of the 1996 HW1 asteroid. (\textit{c})  $k$ = 22 for the equilibrium point $L_D$ ($E_2$) of the M1 Phobos. (\textit{d}) $k$ = 22 for the equilibrium point $L_C$ ($E_r$) of the M1 Phobos. \textit{e}) $k$ = 3.15 for the equilibrium point $L_D$ ($E_2$) of the 243 Ida and 433 Eros asteroids. 	(\textit{f}) $k$ = 3.15 for the equilibrium point $L_C$ ($E_4$) of the 243 Ida and 433 Eros asteroids.}
	\label{asteroids}
\end{figure*}

Figure \ref{asteroids} shows $\mu^*$ versus $\Phi$ and illustrates the stability regions for equilibrum points $L_C$ and $L_D$ for asteroids 1996 HW1, 243 Ida and 433 Eros and M1 Phobos. 

Figure \ref{asteroids} a) and b) plot $\Phi$ $vs.$ $\mu^*$  (27.43, 0.44) for the asteroid 1996 HW1. We observe that the point is outside the region that allows the stability of the equilibrium points $ E_2 $ and $ E_4 $, showing that these equilibrium points are unstable, a result that coincides with the results obtained by \citet{2014Ap&SS.353..105W}

Figures \ref{asteroids} c) and d) show the stability region of the equilibrium points $E_2$ and $E_4$ when $k$ = 22. We plotted the ordered pair (56.09, 0.396) for the M1 Phobos. Due to the characteristics (shape, density and rotation) of M1 Phobos, the equilibrium points $E_2$ and $E_4$ are within the stability region, making these equilibrum points linearly stable. 

The stability of the equilibrium points depends on the bulk density, the shapes, and the angular velocities of the asteroids. The bulk density is obtained from the composition of the asteroid, a characteristic that is hard to change. The shapes of the asteroids are shaped in the long-term in space. On the other hand, the angular velocities of asteroids are altered due to the accelerations caused by the YORP effect \citep{1969JGR....74.4379P}.

\clearpage

Observe that the equilibrium points $E_2$ and $E_4$ of M1 Phobos are close to the boundary that guarantees the condition of stability (see Fig. \ref{asteroids} c and d). If the angular velocity of this body increases, as predicted by the YORP effect, $k$ will decrease, making the equilibrium point to be unstable. This result shows the importance of carrying out a generalized analysis with the aim of globally understanding the dynamic properties in the vicinity of celestial bodies.

Finally, Fig. \ref{asteroids} e) and f) provide information regarding the stability condition for 243 Ida and 433 Eros asteroids. In Table \ref{parameterss} we see that $k$ for asteroids 243 Ida and 433 Eros are very close. Because of this, we will show the results for these two asteroids on the same graph, in Figures \ref{asteroids} e) and f). We plotted ($\phi$, $\mu^*$) = (18.95, 0.26) and ($\phi$, $\mu^*$) = (19.94, 0.23) for 433 Eros and 243 Ida asteroids, respectively.

We observed that the equilibrium points $ E_2 $ and $ E_4 $ (Fig. \ref{asteroids} e) and f), respectively) of asteroids 243 Ida and 433 Eros are unstable due to their physical and dynamical characteristics.

These results show that our generalized analysis coincides with the results obtained for a given asteroid that can be modeled as a rotating mass tripole.

\section{Conclusion}
\label{Conclusion}

Dynamic properties of the rotating mass tripole were addressed in this article. The rotating mass tripole consists of three point masses whose geometric shape depends on the shape of the asteroids under analysis.

We observed that the gravitational potential depends on three free parameters, which are: the force ratio, the mass ratio and the azimuthal angle. We note that the amount of equilibrium points that arise depends on the combination of these free parameters, it can be found from five to eight equilibrium points. The tendency to vary the location of the equilibrium points according to the free parameters is determined.

We also analyzed the topological structure of the zero velocity curves with respect of the azimuthal angle. We observed that the zero velocity curves around the rotating mass tripole have significant changes due to the arched shape of the asteroid.

Analyzing the linearized equations, we observed that the condition of stability of the equilibrium points in region C and D depends of $k$, $\mu^*$ e $\Phi$. For region C, we observed the appearance of bifurcations when $k$ $\textgreater$ 1. On the other hand, the stability of the equilibrium points in region D has two stability transition limits for any value of $k$. For both regions (C and D), it was observed that, as we increase the value of $ k $, the region of stability becomes larger.

Understanding the dynamics of a particle that is subject to the gravitational field of an elongated asteroid is extremely important for the exploration of these bodies. The results presented here provided a global characterization of the dynamic behavior of an infinitesimal mass body around an asteroid modeled as a rotating mass tripole. This allowed a better understanding of the main factors that influence of the topological structure of the gravitational field in the vicinity of asteroids that have an arched shape. More complex models, such as the polyhedral method, are much more accurate and are widely used in the analysis of a specific asteroid, but the present model proved to be useful in providing general information about families of asteroids similar to the tripole model.

\section{Acknowledgements}

The authors wish to express their appreciation for the support provided by: grants 406841/2016-0, 140501/2017-7, 150678/2019-3, 422282/2018-625 9 and 301338/2016-7 from the National Council for Scientific and Technological Development (CNPq); grants 2016/24561-0, 2018/00059-9 and 2016/18418-0, from S\~ao Paulo Research Foundation (FAPESP); grant 88887.374148/2019-00 from the National Council for the Improvement of Higher Education (CAPES) and to the National Institute for Space Research (INPE).

\end{document}